\documentclass[prl,aps]{revtex4}
\usepackage{graphicx}

\begin{document}

\title[Cancellation of ultra-violet  infinities in one loop gravity]{Cancellation of ultra-violet 
 infinities in one loop gravity}
 {\bf  Diploma, 1974}

\author{V.\ E.\ Korepin}
\affiliation{C.N.\ Yang Institute for Theoretical Physics, State
 University of New York at Stony Brook, Stony
Brook, NY 11794-3840}
\date{\today}

\begin{abstract}
This is a historical note.  In 1974 I was an undergraduate  student of L.D. Faddeev.
I was working on quantum gravity [without matter]  in one loop approximation.
I discovered [simultaneously   with G. t'Hooft M. Veltman] that on mass shell
ultra-violet divergences cancel. The text below  is a  translation of the Diploma.

\end{abstract}


\maketitle

The diploma was  in Russian, it  was  never published.
 Recently Faddeev found a reference to my result by an editor  in  Feynman's lectures on 
gravitation \cite{hat}  (the reference is on page xxxvii of the book). The scan of original diploma can be found in the first line of the webpage \cite{korw}. 
The result was obtained simultaneously with G. t'Hooft M. Veltman, see \cite{thv} .
The translation of the diploma follows \footnote{Fictitious particles mentioned in the Diploma now called ghosts}.

\section{Introduction}
The diploma is devoted to quantization of gravity.   First appropriate method for  quantization of gravity was formulated by Dirac  \cite{dir} in 1958 in a frame of Hamiltonian approach. Other methods [essentially equivalent to Dirac's] were suggested in \cite{adm,schw,berg,and,ld}. Hamiltonian approach is not covariant, this makes perturbation theory complicated. 

An approach to covariant quantization of gravity was suggested in \cite{gup}, but this method led to violation of unitarity, see  \cite{fey}. In the same publication Feynman showed that to restore unitarity of a diagram in a form of a closed ring one has to substract another diagram also in a form of a ring, which describes propagation of fictitious particle.   Solution of the problem for any diagram was formulated in 1967 by Fadeev and Popov \cite{pop} and De-Witt \cite{dew} [using essentially different approaches].
This approach to quantization contains fictitious particle [ghosts] and their interaction with gravitons. 
Next question is renormalization. Formally the whole theory is not renormalizable. Here the pure gravity is studied in one loop approximation.

 \section{Covariant Quantization}
 We shall use functional integral for quantization of gravity. The background formulation for generating functional for scattering matrix will be used, not generating functional of Green functions. The integral will depend on asymptotic fields.
 We shall consider scattering matrix directly on mass shell, using classical equation of motion. Matrix element of scattering matrix with $n$ external lines can be obtained by $n$ multiple differentiation of our generating functional with respect to  asymptotic fields. 
 Asymptotic fields are classical fields. The generating functional of scattering matrix  covariantly  depends on asymptotic fields. 
 We will see that some divergencies disappear on mass shell. 
 
 Let us denote by $g_{\mu \nu}^{cl}$ a classical solution of Einstein equations \footnote{In case of upper indices we shall write
 $g^{\mu \nu}_{cl}$ }. So $R=0$ and $R_{\mu \nu}=0$.
 
 Let $iW(g_{\mu \nu}^{cl})$ denote a generating functional for connected Feynman  diagrams with loops:
 \begin{equation} \label{gf}
 e^{iW}= e^{-iS_{Gr}(g_{\rho \omega}^{cl})} \int \exp{iS_{Gr}(g_{\rho \omega}^{cl}+u_{\rho \omega})} \det^{-5/2} \left( -g^{cl}_{\gamma \delta} -u_{\gamma \delta} \right) d^{10} u_{\alpha \beta}
 \end{equation}
 Here 
 \begin{equation} \label{action}
 S_{Gr}(g_{\alpha \beta}) = \int \sqrt {g} R(g_{\alpha \beta})d^{4} x
  \end{equation}
  is Einstein action for gravitational field.  The measure $g^{-5/2} d^{10} g_{\mu \nu}$ assures unitarity. 
  The formula (\ref{gf}) formally covariant under non-abelian group of coordinate transformation. 
  The theory is similar to nonabelian gauge field. We shall use Faddeev-Popov approach to quantization, also we should choose a gauge to keep explicit covariance of $W$. Below we shall use notations:
  \begin{equation} \label{not}
  g^{t}_{\mu \nu}= g^{cl}_{\mu \nu} + u_{\mu \nu}, \qquad \qquad  g_{t}^{\alpha \beta }g^{t}_{\beta \mu}=\delta^{\alpha}_{\mu }
  \end{equation}
  Also $\nabla^{cl} $ will be a covariant derivative with respect to $g^{cl}_{\mu \nu}$ (sometimes we shall write $cl$ as a lower  index ) and 
  
  $\nabla^{t} $ will be a covariant derivative with respect to $g^{t}_{\mu \nu}$ (sometimes we shall write $t$ as a lower  index ) .
  
  Let us introduce an auxiliary condition :
  \begin{equation} \label{ac}
  \nabla_{cl}^{\mu}  U_{\mu \nu}  -{\frac{1}{2}} g^{\alpha \beta}_{cl} {\nabla}_{\nu}^{cl} U_{\alpha \beta} -C_{\nu}=0
   \end{equation}
  Here $C_{nu}$ is an arbitrary function of coordinates. We shall also use
  \begin{equation} \label{der}
D_{\nu}= \nabla^{\mu}_{cl} U_{\mu \nu}  -{\frac{1}{2}} g^{\alpha \beta}_{cl} {\nabla}_{\nu}^{cl} U_{\alpha \beta}  
    \end{equation}
So in one loop quantization we  get :
\begin{equation} \label{ol}
e^{iW}=\int \exp{iS_{Gr}(g^{t}_{\mu \nu})} \det \hat{M} \prod_{x} \delta^{4} \left( D_{\nu}-C_{\nu} \right) g_{t}^{-5/2} d^{10}U_{\mu \nu}
 \end{equation}
 We used $S_{Gr} (g^{cl}_{\alpha \beta})=0$.  The one loop operator  is
 \begin{equation} \label{oper}
 \hat{M}_{\nu \omega}= \nabla_{cl}^{\mu} \nabla_{\mu}^{t} g^{t}_{\nu \omega} +\nabla^{\mu}_{cl}\nabla^{t}_{\nu} g^{t}_{\mu \omega}
 -g_{cl}^{\alpha \beta} \nabla^{cl}_{\nu} \nabla^{t}_{\alpha} g^{t}_{\beta \omega}
  \end{equation}
 It is checked in the original diploma that the right hand side of (\ref{ol}) transforms as scalar density under  coordinate transformations. At this point we want to emphasize that $g^{cl}_{\mu \nu}$ is an arbitrary solution of Einstein equations.
 Let us get rid of  the $\delta$ function following ideas of t'Hooft, see \cite{hooft}.
 Note that the expression 
 \begin{equation} \label{formal}
 N=\left( \prod_{x}\sqrt{g_{cl}} \right) \int dC_{\nu} \exp{\{-i \int d^{4}x \sqrt{g^{cl}} g_{cl}^{\alpha \beta} C_{\alpha} C_{\beta}\}}
  \end{equation}
  does not depend on $g^{cl}_{\mu \nu}$. The constant factor in the equation (\ref{ol}) does not matter, so we can multiply the right hand side by $N$.
  \begin{equation} \label{delta}
e^{iW}=\left( \prod_{x}\sqrt{g_{cl}} \right) \int dC_{\nu} \exp{\{-i \int d^{4}x \sqrt{g^{cl}} g_{cl}^{\alpha \beta} C_{\alpha} C_{\beta}\}}\int \exp{iS_{Gr}(g^{t}_{\mu \nu})} \det \hat{M} \prod_{x} \delta^{4} \left( D_{\nu}-C_{\nu} \right) g_{t}^{-5/2} d^{10}U_{\mu \nu}
 \end{equation}
  After integrating with respect to $C_{\nu}$ we obtain:
  \begin{equation} \label{ghost} 
 e^{iW}=\left( \prod_{x}\sqrt{g_{cl}} \right) \int \exp \{{iS_{Gr}(g^{t}_{\mu \nu})-i \int d^{4}x \sqrt{g^{cl}} g^{\alpha \beta}_{cl}  D_{\alpha}D_{\beta}\}} \left( \det \hat{M} \right) g_{t}^{-5/2} d^{10}U_{\mu \nu}
   \end{equation}
   We can also represent $\det \hat M$ as an integral with respect to anti-commuting vector fields $\overline\chi ^{\alpha}$
  \begin{equation} \label{det}  
   \det \hat M = \int d^{4} \overline\chi ^{\alpha} d^{4} \chi^{\beta} \exp i \int d^{4 }x \overline\chi ^{\alpha} M_{\alpha \beta} \chi^{\beta} 
    \end{equation}
   Finally we arrive to the following expression for generating functional for scattering matrix:
    \begin{equation} \label{fin}  
e^{iW}=\left( \prod_{x}\sqrt{g_{cl}} \right) \int \exp \{{iS_{Gr}(g^{t}_{\mu \nu})-i \int d^{4}x \sqrt{g^{cl}} g^{\alpha \beta}_{cl}  D_{\alpha}D_{\beta}\}} \exp \left(i \int d^{4}x \overline{\chi^{\alpha}} \hat{M}_{\alpha \beta} \chi^{\beta} \right) g_{t}^{-5/2} d^{10}U_{\mu \nu} d^{4}\overline{\chi^{\alpha}} d^{4}\chi^{\beta}
 \end{equation}
  We do not care about common factors.  We can calculate the integral by stationary phase approximation. The stationary point is
   \begin{equation} \label{stap}  
   g_{\mu \nu}= g_{\mu \nu}^{cl}, \qquad \overline{\chi^{\alpha}}=0, \qquad \chi^{\beta} =0
   \end{equation}
 We shall leave in the exponent only terms quadratic in integration variables. In this approximation we get:
   \begin{equation} \label{appr}
    \overline{\chi^{\alpha}} \hat{M}_{\alpha \beta} \chi^{\beta} =  \overline{\chi^{\alpha}} g^{cl}_{\alpha \beta} \nabla^{\mu}_{cl} \nabla^{cl}_{\mu} \chi^{\beta}
     \end{equation}
  The integral
   \begin{eqnarray}  \label{integral}
 & e^{iW}=\\
 & \left( \prod_{x} g_{cl}^{-1} \right) \int \exp \{{i U_{\mu \nu}\frac{\delta^{2}S}{\delta g^{cl}_{\mu \nu} \delta g^{cl}_{\alpha \beta}}U_{\alpha \beta}-i \int d^{4^{\mu}}x \sqrt{g^{cl}} g^{\alpha \beta}_{cl}  D_{\alpha}D_{\beta}\}} \exp \left(i \int d^{4}x \overline{\chi^{\alpha}} \nabla_{cl}^{\mu} \nabla_{\mu}^{cl}\chi^{\alpha} \right)  d^{10}U_{\mu \nu} d^{4}\overline{\chi^{\alpha}} d^{4}\chi^{\beta}  \nonumber
     \end{eqnarray}
  can be taken.   Now we have to calculate the quadratic form:
  \begin{equation}  \label{quadratic}
   U_{\mu \nu}\frac{\delta^{2}S}{\delta g^{cl}_{\mu \nu} \delta g^{cl}_{\alpha \beta}}U_{\alpha \beta}- \int d^{4^{\mu}}x \sqrt{g^{cl}} g^{\alpha \beta}_{cl}  D_{\alpha}D_{\beta} = \frac{1}{2}\int d^{4}x   U_{\mu \nu} \hat{F^{\mu \nu}_{\alpha \beta}}_{cl} g_{cl}^{\alpha \lambda} g_{cl}^{\beta \delta}U_{\lambda \delta}
    \end{equation} 
  In the rest of diploma  we shall use only $g^{cl}_{\alpha \beta}$ , so we shall drop index $cl$. For $\hat{F^{\mu \nu}_{\alpha \beta}}_{cl}$ we get:
  \begin{equation}  \label{eff}
   \hat{F^{\mu \nu}_{\alpha \beta}}_{cl}=\frac{1}{2} \left(\delta_{\rho}^{\mu} \delta^{\nu}_{\lambda} + \delta_{\lambda}^{\mu} \delta^{\nu}_{\rho} - g^{\mu \nu} g_{\rho \lambda} \right) \left(\nabla^{\theta} \nabla_{\theta} \delta^{\rho}_{\alpha} \delta^{\lambda}_{\beta} -2{R^{\rho}_{\alpha}  }^{\lambda}_{\beta} \right)
 \end{equation}   
 It is convenient to denote:
 \begin{equation}  \label{doubleeff}
 \hat{F}_{v}=({\nabla^{\theta} \nabla_{\theta}})_{(v)}
 \end{equation} 
 We can evaluate Gaussian integrals in the form (\ref{integral})
  \begin{equation}  \label{evaluation}
  e^{iW}=\left( \prod_{x} g_{cl}^{-1} \right) \left( \det^{-1/2} \hat{F}_{Gr} \right) \det \hat{F}_{f}
  \end{equation} 
  This is the expression for generating functional of scattering matrix in one loop approximation.  
  
 \section{Cancelation of infinities in one loop approximation}
 To calculate determinants in formula (\ref{evaluation}) we shall use method of proper time
\begin{equation}  \label{proper}
\ln \det F = -\mbox{Tr} \int_{0}^{\infty}\frac{ds}{s}\left(e^{i(F+i0)}-e^{isI} \right)
\end{equation} 
We are going to differentiate the left hand side, so $\mbox{Tr}\int_{0}^{\infty} \frac{ds}{s}e^{isI} $ will not contribute and we shall not write it.

Let us write differential equation and initial data:
\begin{equation}  \label{diff}
\frac{\partial e^{i\hat{F}s}}{\partial s}= i\hat{F} e^{i\hat{F}s}, \qquad e^{i\hat{F}s}|_{s=0}=I
\end{equation} 
Let us denote by $G(x,y|s)$ the kernel of the integral operator $\exp ( {i\hat{F}s})$. 
The main part of the operator $\hat{F}$  is d'Alembertian. So we can separate a singular factor characteristic for parabolic equation:
\begin{equation}  \label{representation}
G(x,x'|s)= \frac{-1}{(4\pi s)^{2}} \exp{\left(\frac{i\sigma(x, x')}{2s}\right)}D^{1/2}A(x,x'|s)
\end{equation} 
Here $A(x,x'|s)$ is a smooth function which turns into $1$ at $s=0$. The $\sigma(x, x')$ is geodesic distance between points $x$ and $x'$. It satisfy a differential equation:
\begin{equation}  \label{geodesic}
g^{\mu \nu} \partial_{\mu} \sigma(x, x') \partial_{\nu} \sigma(x, x') =2 \sigma(x, x')
\end{equation} 
In case of flat space $2 \sigma(x, x') =(x-x')^{2}$. The third factor in (\ref{representation}) is density $D(x,x')=\det  (-\sigma_{\mu \nu'}(x,x')) $.  It satisfy a differential equation:
\begin{equation}  \label{deneq}
D^{-1}\left(\sigma_{.}^{\mu}D \right)_{.\mu}=4
\end{equation} 
It is also convenient to introduce a scalar:
$$
\Delta =\frac{D}{\sqrt{g} \sqrt{g'}}
$$
We use the following notation: $\partial_{\nu}\Phi =\partial \Phi / \partial x^{\nu}$ and $\partial_{\nu '}\Phi =\partial \Phi / \partial {x'}^{\nu}$. Later we shall use $\lim_{x\rightarrow x'}D=g(x) $. This follows from $\lim_{x\rightarrow x'}\partial_{\mu} \partial_{\nu '} \sigma (x, x')=g_{\mu \nu '}(x) $.  In order to calculate $\exp i\hat{F}s$ we need to introduce a function of parallel transport $g^{\alpha}_{\beta'} (x, x')$. It is a by-vector: index $\alpha $ is related to the point $x$ and $\beta'$ to $x'$. The function satisfy the equation: $\sigma_{.}^{\tau} g^{\alpha}_{\beta'  \tau} = 0 $. Corresponding boundary condition is $g^{\alpha}_{\beta'} (x, x') \rightarrow \delta^{\alpha}_{\beta'}$ as $x\rightarrow x'$ \footnote{In flat space $g^{\alpha}_{\beta'} (x, x')=  \delta^{\alpha}_{\beta'}$ }.
The function of parallel transport  has the following properties:
$$ 
g_{\mu \nu'}=g_{\nu' \mu}, \quad g_{\mu}^{\nu'}\sigma_{. \nu'}=- \sigma_{. \mu}, \quad g_{\mu \sigma'} g_{\nu}^{\sigma'}= g_{\mu \nu}, \quad  \det \left(-g_{\mu \nu'} \right) =\sqrt{g g'}
$$
Now we ready to study the formula (\ref{representation}).  For fictitious particles we put
\begin{equation}  \label{substitute}
A(x,y|s)^{\alpha}_{\beta'}=g^{\alpha}_{\beta'} (x, y) f^{f}(x,y|s)
\end{equation} 
The function $ f^{f}(x,y|s)$ is a by-scalar satisfying equation:
\begin{equation}  \label{auxiliary}
\frac{\partial f^{f}}{\partial s} + \frac{\sigma_{.}{\mu}f^{f}_{.\mu}}{s} = \frac{i}{4} g^{\alpha}_{\beta '} \Delta ^{-1/2} \nabla^{\mu}\nabla_{\mu} \left( \Delta^{1/2} g_{\alpha}^{\beta'} f^{f} \right)
\end{equation} 
Consider Taylor series:
\begin{equation}  \label{auxiliaryser}
f^{f}(x,y|s)=\sum_{n=0}^{\infty}a_{n}^{f}(x,y) (is)^{n}, \qquad a_{0}=1
\end{equation}
Coefficients satisfy equations:
\begin{equation}  \label{recursion}
\sigma_{.}^{\mu}a_{n . \mu} + na_{n} = \frac{1}{4} \Delta^{-1/2} g^{\alpha}_{\beta'} \left(g_{\alpha}^{\beta'}\Delta^{1/2} a_{n-1} \right)_{. \theta}^{\theta}
\end{equation}
Let us do similar calculations for gravitons:
\begin{equation}  \label{graviton}
A(x,y|s)^{\mu \nu}_{\alpha' \beta '}=\frac{1}{2} \left(g^{\mu}_{\lambda'}g^{\nu}_{\gamma'}+g^{\mu}_{\gamma'} g^{\nu}_{\lambda'}-\frac{1}{2}g^{\mu \nu}g_{\gamma' \lambda'} \right) f^{\gamma' \lambda'}_{\alpha' \beta'}(x,y|s)
\end{equation}
Here $f$ is a scalar at $x$ and 4-tensor at point $y$, it is symmetric and traceless with respect to $\gamma' \lambda'$.
It satisfy an equation:
\begin{equation}  \label{grecursion}
\frac{\partial}{\partial s } f^{\mu' \nu'}_{\alpha' \beta'} + \frac{\sigma_{.}^{\omega}}{s} (f_{.\omega})^{\mu' \nu'}_{\alpha' \beta'}= i g_{\omega}^{\mu'}g_{\delta}^{\nu'} \Delta^{-1/2}\left(\Delta^{1/2}g^{\omega}_{\lambda'}g^{\delta}_{\nu'}f^{\lambda' \nu'}_{\alpha' \beta'} \right)_{.\theta}^{\theta} - 2i g_{\theta}^{\mu'}g^{\nu'}_{\varsigma}{R^{\theta}_{\omega}}^{\varsigma}_{\delta} g^{\omega}_{\lambda'}g^{\delta}_{\gamma'} f^{\lambda' \gamma'}_{\alpha' \beta'}
\end{equation}
Consider Taylor series for this $f$
\begin{equation}  \label{gtaylor}
f^{\mu' \nu'}_{\alpha' \beta'}=\sum_{n=0}^{\infty}a^{\mu' \nu'}_{\alpha' \beta'} (is)^{n}
\end{equation}
Coefficients satisfy a recursion: 
\begin{equation}  \label{gravirecursion}
 \sigma_{.}^{\omega}(a_{n.\omega})^{\mu' \nu'}_{\alpha' \beta'} + n(a_{n})^{\mu' \nu'}_{\alpha' \beta'}= g_{\omega}^{\mu'}g_{\delta}^{\nu'} \Delta^{-1/2}\left(\Delta^{1/2}g^{\omega}_{\lambda'}g^{\delta}_{\nu'}(a_{n-1})^{\lambda' \nu'}_{\alpha' \beta'} \right)_{.\theta}^{\theta} - 2g_{\theta}^{\mu'}g^{\nu'}_{\varsigma}{R^{\theta}_{\omega}}^{\varsigma}_{\delta} g^{\omega}_{\lambda'}g^{\delta}_{\gamma'} (a_{n-1})^{\lambda' \gamma'}_{\alpha' \beta'}
\end{equation}
Note that the coefficients $(a_{n})^{\lambda' \gamma'}_{\alpha' \beta'}$  are symmetric and traceless with respect to upper indices 
${\lambda' \gamma'}$. Also 
\begin{equation}  \label{foundation}
(a_{0})^{\lambda' \gamma'}_{\alpha' \beta'}=\frac{1}{2} \left(\delta^{\lambda'}_{\alpha'}\delta^{\gamma'}_{\beta'}+ \delta^{\lambda'}_{\beta'}\delta^{\gamma'}_{\alpha'}  - \frac{1}{2}g^{\lambda' \gamma'} g_{\alpha' \beta'} \right)
\end{equation}
These calculations directly generalize the ones by B.S. De-Witt [he used them for description of interaction of scalar particles with external gravity].  So we described the kernel  $\exp{(is\hat{F})}$ , see (\ref{representation}) . We can use it to separate infinities in the formula (\ref{proper}).  Ultraviolet infinities arise from integration at $s\sim 0$. Taylor expansion (\ref{auxiliary}) and (\ref{gtaylor}) are useful.  For fictitious particles we obtain:
 \begin{equation}  \label{ficdet}
\ln \det \hat{F}^{f} =4\int dx^{4} \sqrt{g}\sum_{n=0}^{\infty}\int_{0}^{\infty}\frac{ds}{s(4\pi s)^{2}}e^{\frac{i\sigma}{2s}} (is)^{n} a_{n}(x,x)
\end{equation}
We put $\exp{\frac{i\sigma}{2s}}|_{0}=1$. Only coefficients at $a_{0}, a_{1}$ and $a_{2}$ are divergent at zero [quartic, square and logarithmic divergencies correspondingly]. So in (\ref{ficdet}) we shall consider only first three terms:
\begin{eqnarray}  
&{\ln \det \hat{F}^{f}} = 4 \int dx^{4} \sqrt{g}\int_{0}^{\infty}\frac{ds}{s(4\pi s)^{2}}e^{\frac{i\sigma}{2s}} (1+isa_{1}-s^{2}a_{2} )=\\ \nonumber
& = 4 \int \frac{d^{4}x \sqrt{g}}{(4\pi)^{2}}\left( \frac{-4}{(\sigma +i0)^{2}}  -\frac{2a_{1}}{\sigma+i0}+[\ln\frac{\sigma +i0}{2} -\int_{0}^{\infty}\frac{ds}{s} e^{i/s}] a_{2}\right)  \label{ficinf}
\end{eqnarray}\
Last integral is divergent at $0$ and $\infty$.  We can get similar expression for gravitons:
\begin{equation}  \label{grainf}
\ln \det \hat{F}^{Gr}=\frac{1}{(4\pi)^{2}}\int d^{4}x \sqrt{g} \mbox{tr} \left(\frac{-4}{(\sigma +i0)^{2}} I  -\frac{2\hat{a}_{1}}{\sigma+i0}+[\ln\frac{\sigma +i0}{2} -\int_{0}^{\infty}\frac{ds}{s} e^{i/s}] \hat{a}_{2}  \right)|_{x=y}
\end{equation}
So we need coefficients $a_{1}$ and $a_{2}$ for both gravitons and fictitious  particles.  Then we can use (\ref{evaluation}) to calculate divergencies of the generating functional for scattering matrix. We can use (\ref{recursion}) to derive:
\begin{equation}  \label{step}
\lim_{x\rightarrow y} a^{f}_{n}=\frac{1}{4n} \lim_{x\rightarrow y}\Delta^{-1/2} g_{\beta'}^{\alpha} \left(g_{\alpha}^{\beta'} \Delta^{1/2}a_{n-1}^{f} \right)_{.\theta}^{\theta}
\end{equation}
Because the 
$
\lim_{x\rightarrow y}\sigma_{.}^{\mu}=0.
$
Equation (\ref{gravirecursion}) lead to:
\begin{equation}  \label{expression}
\lim  (a_{n})^{\mu' \nu'}_{\alpha' \beta'} = \frac{1}{n}\lim_{x\rightarrow y} g_{\omega}^{\mu'}g_{\delta}^{\nu'} \left(\Delta^{1/2}g^{\omega}_{\lambda'}g^{\delta}_{\nu'}(a_{n-1})^{\lambda' \nu'}_{\alpha' \beta'} \right)_{.\theta}^{\theta} - 2{R^{\mu'}_{\lambda'}}^{\nu'}_{\gamma'}  (a_{n-1})^{\lambda' \gamma'}_{\alpha' \beta'}
\end{equation}
In order to calculate the right hand side in these equations we need to know expressions like $\sigma_{.\mu \nu \gamma \delta}$, which we can find recursively  from equations: $\sigma_{.}^{\mu} \sigma_{.\mu}=2\sigma  $ also  $\sigma_{.}^{\mu} g^{\alpha}_{\beta' \mu}=0$ and $4\Delta^{1/2}=2{\Delta^{1/2}}_{.}^{\mu}\sigma_{.\mu} +\Delta^{1/2} \sigma_{.\mu}^{\mu}$ .
Last equation follows from (\ref{deneq}). Now shall evaluate covariant derivatives and use commutation rule:
\begin{equation}  \label{commutation}
\left(\phi ^{\mu} \right)_{.\nu \sigma} -\left(\phi ^{\mu} \right)_{. \sigma \nu }= {R_{\nu \sigma}}^{\mu}_{\tau} \phi^{\tau}
\end{equation}
Let us present a table of limits:
\begin{eqnarray}  \label{table}
&\lim \sigma =\lim \sigma_{.}^{\mu}=0, \qquad \lim \sigma_{.\mu \nu} =g_{\mu \nu}, \qquad \lim \sigma_{.\alpha \beta \gamma}=0 \qquad  \lim \sigma_{.\nu \sigma \tau \rho} =\frac{1}{3}(R_{\nu \tau \sigma \rho}+ R_{\nu \rho \sigma \tau}) \\
&\lim {{\sigma_{.\mu}^{\mu}}_{\nu}^{\nu}}_{\sigma}^{\sigma}=\frac{8}{5}R_{.\mu}^{\mu}+ \frac{4}{15}R_{\mu \nu}R^{\mu \nu}-\frac{4}{15}R_{\alpha \beta \gamma \delta}R^{\alpha \beta \gamma \delta}\\
& \lim\Delta^{1/2}=1, \qquad \lim\left( \Delta^{1/2}\right)_{.\mu}=0, \qquad  \lim\Delta^{1/2}_{.\mu \nu}=-\frac{1}{6}R_{\mu \nu},  \qquad 
 \lim \left( \Delta^{1/2}\right)_{.\mu \nu}^{\nu}=-\frac{1}{6}R_{.\mu } \\
 &  \lim \left( \Delta^{1/2}\right)_{.\mu \nu}^{\mu \nu}=-\frac{1}{5}R_{.\mu }^{\mu} +\frac{1}{36}R^{2}-\frac{1}{30}R_{\mu \nu}R^{\mu \nu} +\frac{1}{30}R_{\alpha \beta \gamma \delta}R^{\alpha \beta \gamma \delta}
\end{eqnarray}
These limits were evaluated by  De Witt in \cite{hooft}.
\begin{eqnarray}  \label{tablaux}
&\lim g_{\nu'}^{\mu}=\delta_{\nu'}^{\mu}, \qquad  \lim g_{\nu' . \tau}^{\mu}=0, \qquad   \lim g_{\nu' . \rho \lambda}^{\mu}=\frac{1}{2} R_{\nu'  \rho \lambda}^{\mu} \\
& \lim g_{\nu' . \rho \lambda \alpha}^{\mu}=\frac{1}{3} \left( R_{\rho \lambda \nu'  .\alpha}^{\mu} +  R_{\rho \lambda \nu'  .\lambda}^{\mu}\right), \qquad \lim g_{\nu' . \rho  \alpha}^{\mu \quad  \rho \alpha }=\frac{1}{2} \left( R_{\alpha \beta \quad \nu'  .}^{\quad \mu \quad \alpha \beta} -  R_{\alpha \beta \gamma}^{\qquad \mu}R^{\alpha \beta \gamma}_{\qquad \nu'} \right)
\end{eqnarray}
Using these tables and formula (\ref{step}) we calculate the coefficients 
\begin{eqnarray}  \label{coeff}
&a_{0}^{f}=1, \qquad  \lim a^{f}_{1}=\frac{1}{4} \delta_{\alpha}^{\beta'} \left(\Delta^{1/2}g^{\alpha}_{\beta'} \right)_{.\mu}^{\mu}=-\frac{1}{6}R=0 \\
&\lim a_{2}^{f}=\frac{1}{32}\lim \delta_{\alpha}^{\beta'}\left[ \Delta^{1/2}g^{\alpha}_{\beta'}g_{\gamma}^{\delta'}\left(g^{\gamma}_{\delta'} \Delta^{1/2}\right)_{.\iota}^{\iota} \right]_{.\mu}^{\mu}=-\frac{1}{10}R_{.\mu}^{\mu}+\frac{1}{72}R^{2}-\frac{1}{60}R_{\mu \nu} R^{\mu \nu} -\frac{11}{240}R_{\alpha \beta \gamma \delta}R^{\alpha \beta \gamma \delta}\\
&\lim a_{2}^{f}= -\frac{11}{240}R_{\alpha \beta \gamma \delta}R^{\alpha \beta \gamma \delta}
\end{eqnarray}
Similar for gravitons:
 \begin{eqnarray}  \label{grcoeff}
\mbox{tr}a_{0}=9, \quad  \mbox{tr}a_{1}=0, \quad  \mbox{tr}a_{2}=\frac{21}{40}R_{\alpha \beta \gamma \delta}R^{\alpha \beta \gamma \delta}
\end{eqnarray}
Quadratic divergencies are absent both for gravitons and fictitious particles: $a_{1}^{f}=0$ and $a_{1}^{Gr}=0$
As for logarithmic divergencies, we should take into account the identity \cite{hooft}:
\begin{equation}  \label{identity}
\int d^{4} x\sqrt{g}\left(R^{2}-4 R_{\mu \nu}R^{\mu \nu} + R_{\alpha \beta \gamma \delta}R^{\alpha \beta \gamma \delta}\right) =0
\end{equation}
This means that 
$$
\int d^{4} x\sqrt{g}\left( R_{\alpha \beta \gamma \delta}R^{\alpha \beta \gamma \delta}\right) =0
$$
This means that logarithmic divergencies also absent. So we proved that there is no ultra-violet divergencies for generating functional of scattering matrix on mass shell. This result was obtained simultaneously with \cite{thv}.

\section{Finite part of the generating functional of the scattering matrix}
We can rewrite equation (\ref{auxiliary}) for Fourier transform:
$$
f^{f}(x,x|s)=\int_{-\infty}^{\infty} d\omega e^{is\omega} f^{f}(x,x|\omega)
$$
\begin{equation}  \label{ficform}
\frac{\partial (\omega f^{f})}{\partial \omega}= \sigma_{.}^{\mu}f^{f}_{.\mu}+\frac{1}{4}g^{\alpha}_{\beta'}\Delta^{-1/2}\left(\nabla^{\mu}\nabla_{\mu} \Delta^{1/2}g_{\alpha}^{\beta'}\frac{\partial f^{f}}{\partial \omega} \right)
\end{equation}
Similar for gravitons we can define Fourier transform of $f^{Gr}$:
$$
f^{\mu' \nu'}_{\alpha' \beta'}(x,x|s)=\int_{-\infty}^{\infty} d\omega e^{is\omega} f^{\mu' \nu'}_{\alpha' \beta'}(x,x|\omega)
$$
We can start from equation (\ref{grecursion}) and obtain:  
\begin{equation}  \label{grform}
\frac{\partial (\omega f^{\mu' \nu'}_{\alpha' \beta'})}{\partial \omega }  = \sigma_{.}^{\lambda} \left(f_{.\lambda}  \right)^{\mu' \nu'}_{\alpha' \beta'}  +  g_{\omega}^{\mu'}g_{\delta}^{\nu'} \Delta^{-1/2}\left(\Delta^{1/2}g^{\omega}_{\lambda'}g^{\delta}_{\nu'}\frac{\partial f^{\lambda' \nu'}_{\alpha' \beta'} }{\partial \omega}\right)_{.\theta}^{\theta} - 2 g_{\theta}^{\mu'}g^{\nu'}_{\varsigma}{R^{\theta}_{\omega}}^{\varsigma}_{\delta} g^{\omega}_{\lambda'}g^{\delta}_{\gamma'} \frac{ \partial f^{\lambda' \gamma'}_{\alpha' \beta'}}{\partial \omega}
\end{equation}
These function are used in diploma \cite{korw} to represent the finite part of  the generating functional of the scattering matrix
\begin{equation}  \label{finitepart}
iW= \frac{1}{2(4\pi)^{2}}\int d^{4}x\sqrt{g}\int_{-\infty}^{\infty}d\omega \{4f^{f}(x,x|\omega)-\frac{1}{2} \mbox{tr}f^{Gr}(x,x|\omega)  \}
\omega^{2}\ln \frac{\omega +i0}{m^{2}}\end{equation}
Here $m^{2}$ is an arbitrary  positive constant. The equations for $f$ should be solved by perturbations starting form the flat metric.
\section{One loop diagram with two vertices }
The section of diploma consider insertions of one loop diagram with one and two vertices in a tree diagram.
The calculations in the diploma proves that the insertion of the diagram with one and two loops in any tree diagram vanish.
The full text of diploma [in Russian] can be found on authors web-page \cite{korw}.

\end{document}